# Properties of galaxies reproduced by a hydrodynamic simulation

M. Vogelsberger[1], S. Genel[2], V. Springel[3,4], P. Torrey[2], D. Sijacki[5], D. Xu[3], G. Snyder[6], S. Bird[7], D. Nelson[2] & L. Hernquist[2]

[1] Department of Physics, Kavli Institute for Astrophysics and Space Research, Massachusetts Institute of Technology, Cambridge, MA 02139, USA
[2] Harvard-Smithsonian Center for Astrophysics, 60 Garden Street, Cambridge, MA 02138, USA
[3] Heidelberg Institute for Theoretical Studies, Schloss-Wolfsbrunnenweg 35, 69118 Heidelberg, Germany
[4] Zentrum für Astronomie der Universität Heidelberg, ARI, Mönchhofstr. 12-14, 69120 Heidelberg, Germany
[5] Kavli Institute for Cosmology, and Institute of Astronomy, Madingley Road, Cambridge, CB3 0HA, UK
[6] Space Telescope Science Institute, 3700 San Martin Drive, Baltimore, MD 21218, USA
[7] Institute for Advanced Study, 1 Einstein Drive, Princeton, NJ, 08540, USA

**Previous simulations of the growth of cosmic structures have broadly reproduced the 'cosmic web' of galaxies that we see in the Universe, but failed to create a mixed population of elliptical and spiral galaxies due to numerical inaccuracies and incomplete physical models. Moreover, because of computational constraints, they were unable to track the small scale evolution of gas and stars to the present epoch within a representative portion of the Universe. Here we report a simulation that starts 12 million years after the Big Bang, and traces 13 billion years of cosmic evolution with 12 billion resolution elements in a volume of $(106.5\ \text{Mpc})^3$. It yields a reasonable population of ellipticals and spirals, reproduces the distribution of galaxies in clusters and statistics of hydrogen on large scales, and at the same time the metal and hydrogen content of galaxies on small scales.**

The initial conditions for structure formation in the Universe are tightly constrained from measurements of anisotropies in the cosmic microwave background radiation[1]. However, previous attempts to reproduce the properties of the observed cosmological structures with computer models have shown only limited success. No single, self-consistent simulation of the Universe was able to simultaneously predict statistics on large scales, such as the distribution of neutral hydrogen or the galaxy population of massive galaxy clusters, together with galaxy properties on small scales, such as the morphology and detailed gas and stellar content of galaxies. The challenge lies in following the baryonic component of the Universe using hydrodynamic simulations[2-4], which are required to model gas, stars, supermassive black holes (SMBHs), and their related energetic feedback. The vast computational challenges of these simulations has forced previous attempts to focus either on simulating only small portions of the Universe or to employ a coarse resolution such that internal characteristics of galaxies could not be resolved. Furthermore, common numerical techniques are either highly spatially adaptive, or very accurate, sacrificing one of those aspects. Finally, poorly-understood small-scale processes such as star-formation and accretion onto SMBHs are coupled to galactic and super-galactic scales, introducing large uncertainties into modelling techniques.

Rapid advances in computing power combined with improved numerical algorithms and more faithful models of the relevant physics have allowed us to produce a simulation ("Illustris") that simultaneously follows the evolution of dark matter and baryons in detail. Starting approximately 12 million years after the Big Bang, our simulation tracks the evolution of more than 12 billion resolution elements in a volume of $(106.5 \text{ Mpc})^3$ up to the current epoch (redshift $z = 0$). This allows us to achieve a dark matter mass resolution of $6.26 \times 10^6 \text{ M}_\odot$, and a baryonic mass resolution of $1.26 \times 10^6 \text{ M}_\odot$. The smallest scale over which the hydrodynamics is resolved is 48 pc, whereas gravitational forces are resolved down to 710 pc at z = 0 and even to smaller scales at high redshifts (e.g., 473 pc at z = 2). Our calculation therefore overcomes the problems of previous hydrodynamic simulations which either did not cover a large enough portion of the Universe to be representative, lacked adequate resolution, or failed to reach the present epoch. Aside from having a large volume and improved resolution, our simulation is evolved with the novel hydrodynamic algorithm AREPO[5], which uses a moving unstructured Voronoi tessellation in combination with a finite volume approach. Finally, we employ a numerically well-posed and reasonably complete model for galaxy formation physics, which includes the formation of both stars and SMBHs, and their effects on their environments in forms of galactic super-winds driven by star-formation, as well as radio bubbles and radiation proximity effects caused by active galactic nuclei (AGNs) (see Methods).

Unlike previous attempts, we find a mix of galaxy morphologies ranging from blue spiral galaxies to red ellipticals, with a hydrogen and metal content in good agreement with observational data. At the same time, our model predicts correctly the large scale distribution of neutral hydrogen, and the radial distribution of satellite galaxies within galaxy clusters. Our results therefore demonstrate, that the ΛCDM model can correctly describe the variety of observational data on small and large scales in our Universe. It also predicts a strong, scale-dependent impact of baryonic effects on the dark matter distribution, at a level that has significant implications for future precision probes of cosmology.

**Observing the model universe**

The simulation volume contains 41, 416 galaxies at z = 0 that are resolved with more than 500 stellar resolution elements. Our model yields a population of non-star-forming elliptical galaxies, star-forming disk galaxies, and irregular galaxies (Fig. 1a). We find that galaxies with low star-formation rates contain about 52% of all the stellar mass that is in galaxies more massive than $M_* = 10^9$ $M_\odot$, which agrees well with observations[6] (54% − 60%). Simulating the formation of realistic disk galaxies, like our own Milky Way, has remained an unsolved problem for more than two decades. The culprit was an angular momentum deficit leading to too high central concentrations, overly massive bulges, and unrealistic rotation curves[7,8]. The fact that our calculation naturally produces a morphological mix of realistic disk galaxies coexisting with a population of ellipticals resolves this long-standing issue. It also shows that previous futile attempts to achieve this were not due to an inherent flaw of the ΛCDM paradigm, but rather due to limitations of numerical algorithms and physical modelling.

As our simulation follows the evolution of galaxies starting shortly after the Big Bang, we can construct virtual mock observations that mimic the conditions that the Hubble Space Telescope encounters as it images galaxies across cosmic time in very deep surveys, such as the Ultra Deep Field (UDF), and the ongoing Frontier Fields program. These observations capture a large variety of galaxy luminosities, sizes, colours, morphologies, and evolutionary stages, providing remarkable benchmarks for galaxy formation theories. We have constructed a mock UDF and compare it side by side to data from the HST eXtreme Deep Field (XDF) compilation[9] (Fig. 1b, 1c). Galaxies in the mock UDF appear strikingly similar to the observed population in terms of number density, colours, sizes and morphologies. Our model is the first hydrodynamic simulation from which a faithful deep UDF-like observation could be constructed, thanks to its combination of large volume and high resolution, along with the new numerical techniques that allow it to reproduce realistic galaxy morphologies.

**Satellite galaxies in clusters**

This qualitative agreement also extends to many quantitative probes of the distribution and internal structure of galaxies. The abundance and spatial distribution of satellite galaxies forms an important observational test of the ΛCDM paradigm, as it is very sensitive both to details of the hierarchical structure formation process and to galactic-scale baryonic processes. In fact, two of the most acute challenges to the ΛCDM model are related to satellites on galactic scales, giving rise to the "missing satellite"[10] and "too-big-to-fail" problems[11]. A particularly taxing problem is the radial distribution of satellite galaxies within galaxy clusters, which are the largest gravitationally bound objects in the Universe.

Recently, large SDSS cluster samples have made it possible to measure accurate radial satellite profiles[12,13]. Theoretical models have so far struggled to reproduce the shapes and normalisations of these profiles. Semi-analytic models have resorted to ad-hoc prescriptions for mass-stripping and tuned approximations for satellite orbits since they are missing the gravitational effects of the stellar component. Despite the freedom offered by the adopted coarse parameterisations, most semi-analytic models consistently find radial profiles that are too steep, and in particular overestimate the number of satellites in the inner regions of clusters (r < 100 - 200 kpc)[13-15]. Previous hydrodynamic simulations, on the other hand, found inner radial profiles that are too shallow and in most cases could not reproduce the observed normalisation[16-18], owing both to limited resolution and to an over-production of stars in satellites due to missing physical processes.

For the most massive haloes, we have calculated a stacked projected galaxy count profile and compared it to results from a sample[13] of satellite galaxies brighter than r < −20.5, for host systems at 0.15 ≤ z ≤ 0.4 extracted from 55,121 SDSS groups and clusters ($10^{13.7}$ $M_\odot$ < $M_{500,crit}$ < $10^{15.0}$ $M_\odot$) centred on luminous red galaxies (Fig. 2). Having higher resolution than previous hydrodynamic simulations, and more realistic feedback models that suppress the stellar masses of satellites, we obtain a good agreement with both the observed profile shape and its normalisation, as well as with the mean galaxy colour as a function of cluster-centric distance[19]. The radial distribution of dark matter subhaloes from a corresponding dark matter only simulation ("Illustris-Dark") flattens towards the centre. This difference is due to dissipational processes of galaxy formation that make the stellar component more resistant to tidal disruption close to cluster centers. This directly demonstrates that neglecting baryonic physics causes inaccuracies in the spatial distribution of satellite galaxies which in turn can lead to errors in estimates of galaxy merger rates.

**Metals and neutral hydrogen in galaxies**

The high mass resolution of the simulation makes it possible to study the internal characteristics of galaxies. It allows us, for example, to make clear predictions for the stellar and gaseous contents of galaxies, and for the baryonic cycle operating between them. Our model predicts the present-day HI content of galaxies in the local Universe, which can be compared to observations as revealed by the Arecibo Legacy FAST ALFA Survey (ALFALFA) (Fig. 3a). As stars form out of cold gas which is largely neutral, the HI richness is a good probe of the reservoir of gas available for star formation in a galaxy. The HI- selected samples of ALFALFA are typically biased towards the most gas-rich star-forming galaxies; i.e. the major limitation of HI-selected samples is that they will miss the most gas-poor elliptical galaxies.

Our results recover the trend of decreasing HI richness with increasing stellar mass. For the most massive galaxies, the full simulated galaxy sample deviates from the observed mean relation. Many of these massive galaxies have been quenched through AGN feedback and therefore contain little amounts of gas; i.e. they would not be detected by ALFALFA given their low HI content. If we focus instead on star-forming galaxies we find significantly better agreement even at higher masses, and we reproduce the observed HI richness relation over nearly four orders of magnitude in stellar mass. We also include a separate HI richness relation for the satellites of the most massive cluster and compare this to HI measurements of galaxies in the Virgo cluster[20] where ram pressure removes HI from infalling galaxies. Such hydrodynamic processes are not directly accessible to semi-analytic models but are captured in our simulation. We can therefore recover the observed trend: satellites in cluster environments have a lower HI content compared to the whole population. The predicted difference is not as pronounced as in the observations, but this may well be caused by the fact that our volume does not contain a galaxy cluster as massive as Virgo.

Every dynamical time, a small fraction of the galactic cold neutral gas turns into stars, which inherit the chemical composition of the gas. Stellar metallicities therefore probe different processes intrinsic to galaxy formation: chemical enrichment, feedback, and in- and outflows. Hydrodynamic cosmological simulations have so far been unable to produce correct stellar metallicities; measures such as the mean stellar metallicity or the cosmic density of total metal mass locked up in stars, $\rho_{Z,*}$, were discrepant from the observed values by almost a factor of two[21,22]. Our model, on the other hand, produces $\rho_{Z,*} = 7.75 \times 10^6\ M_\odot\ /\ Mpc^3$, in agreement with the observed value[23] of $\rho_{Z,*} = (7.1 \pm 2) \times 10^6\ M_\odot\ /\ Mpc^3$.

Recent observations find a relation where more massive galaxies contain a larger proportion of metals[24-26]. Our predictions agree well with these observations, and, most importantly, the simulation recovers the flattening of the relation above $M \sim 10^{11}\ M_\odot$ (Fig. 3b). The observed trend is reproduced over nearly five orders of magnitude in stellar mass. These agreements with observations are driven by the combination of our accurate hydrodynamic scheme together with realistic effective feedback models and a complete stellar evolution prescription, a combination that was lacking in previous studies.

**Large-scale characteristics of neutral hydrogen**

The space between galaxies, the intergalactic medium, is filled with low-density, warm gas, mainly comprised of ionised hydrogen. Observationally, the intergalactic medium can be probed through the forest of Lyman-α absorption lines in quasar spectra. The corresponding distribution of hydrogen column densities is well-constrained[27,28] over a wide dynamic range. At high gas densities, hydrogen becomes self-shielded from the ionising UV background and forms dense neutral clouds with a distinctive spectral absorption signature, the so-called damped Lyman-α absorbers (DLAs), which are observed mostly at redshift z ∼ 2 - 5 and probe initial stages of galaxy formation.

Early hydrodynamic simulations successfully predicted the low column density statistics of the Lyman-α forest[29], whereas describing the main properties of higher density absorbers and DLAs correctly only became feasible over the last few years[30,31]. However, recent numerical calculations have failed to explain the metallicity distribution of DLAs, which is now reasonably well constrained by data[32]. The outcomes of most simulations gave metallicities for DLAs that are too high or yielded a metallicity distribution that is too broad, in disagreement with observations[33,34].

We contrast our predicted HI column density distribution function (CDDF) (Fig. 4a) and the metal content of DLAs (Fig. 4b) with observations. For the DLA metallicities, we compare to an observational compendium[32], based on all available quasars between z = 2 and z = 4, while the CDDF data is centered at z = 3. The properties of DLAs are sensitive to the balance between gas accretion, outflows and ionisation. Probing such a complex interplay of processes is a particular strength of hydrodynamic simulations. The prediction of our model is in remarkable agreement with the observational data, reproducing in detail the shape and location of the metallicity peak. This success is a result of accurate hydrodynamics and modeling of galactic super-winds resulting in metal outflows. Also, we find good agreement between our theoretical predictions and the observed CDDF.

At low redshift, most of the baryons have not yet been detected. Improved instrumental sensitivity may lead to their detection in the foreseeable future[35], at a level depending on exactly where they reside, and at what temperature, which is a matter of theoretical debate. Based on our model we predict that, at z = 0, gas that is not bound to any haloes constitutes 81% of the total baryons in the Universe, but contains only 34% of the heavy metals.

**The impact of baryons on dark matter**

The discussion thus far has stressed that a direct modelling of baryonic physics is essential to soundly connect cosmological predictions to galaxy observations. Even beyond that, baryonic processes can actually impact and modify the dark matter distribution, and therefore alter the matter power spectrum *P(k)*. Measuring *P(k)* provides a powerful cosmological probe since theoretical models can be used to connect the present-day *P(k)* to the initial power spectrum. Such measurements come, for example, from weak lensing studies, galaxy clustering surveys, and the analysis of the Lyman-α forest. However, the interpretation of upcoming weak lensing surveys such as EUCLID, that will measure *P(k)* on 0.1 $h$ Mpc$^{-1}$ < $k$ < 10 $h$ Mpc$^{-1}$ scales, must consider the impact of baryons in order to achieve the required level of ~1% accuracy[36,37]. This effect has often been neglected, even though it is not restricted to just the smallest spatial scales.

We have calculated the dimensionless matter power spectrum, $\Delta^2(k) = k^3 P(k)/(2\pi^2)$, along with the corresponding dark matter-only result (Fig. 5). The dynamic range of our simulation allows us to probe *P(k)* on a wider range of scales than ever possible before through a single hydrodynamic simulation. On scales smaller than $k$ ~ 1$h$ Mpc$^{-1}$, AGN-driven outflows reduce the total power in a scale-dependent way by up to ∼ 30 - 40% compared to the dark matter-only prediction. This impact is even larger than found in previous studies[38], which is most likely related to the strong radio-mode AGN feedback in our simulation needed to match the stellar mass content of massive haloes. On smaller scales, gas cooling processes become important and enhance the power on scales smaller than $k$ ~ 100 $h$ Mpc$^{-1}$. Here, the power spectrum can deviate by a factor of a few compared to collisionless results. Measuring this fundamental statistic precisely hence requires high-resolution hydrodynamic simulations in large volumes; it cannot be done by dark matter-only simulations or semi-analytic modelling. We also present the predictions of commonly employed empirical fitting models for the non-linear evolution[39,40] of *P(k)*. Those models have been calibrated on dark matter-only simulations and clearly fail to describe the results of full hydrodynamic calculations, rendering their use impractical for the precision requirements of upcoming surveys.

**Looking ahead**

Although our simulation provides a significant step forward in modelling galaxy formation by reproducing simultaneously many disparate observations on large and small scales, there are still outstanding problems. One such problem lies in the formation of low mass galaxies: Our simulation tends to build up the stellar mass of low mass galaxies below $M_* \sim 10^{10}$ $M_\odot$ too early, resulting in stellar populations that are too old with mean ages a factor two to three larger than observed. This tension is shared by semi-analytic models and other recent hydrodynamic simulations alike, pointing towards an open problem in low-mass galaxy formation[41]. It remains to be seen whether new stellar feedback models that for example include effects of radiation pressure on dust can resolve this issue[42]. It will clearly be challenging to test new schemes that also directly treat the stellar radiation fields with statistically meaningful samples of galaxies, as this poses extremely high computational demands that go significantly beyond what was achieved in the present work. Nevertheless, such new generations of large-scale high-resolution hydrodynamic simulations might become feasible within the next decade.

**Methods Summary**

The equations of gravity and hydrodynamics are evolved using the moving-mesh code AREPO[5] combined with a galaxy formation model which includes[43]: gas cooling with radiative self-shielding corrections, star formation, energetic feedback from growing SMBHs and exploding supernovae, stellar evolution with associated chemical enrichment and stellar mass loss, and radiation proximity effects for AGNs. Our AGN feedback consists of three components: thermal quasar-mode feedback, thermal-mechanical radio-mode feedback, and radiative feedback. The efficiency of physical processes below the resolution scale are treated in parameterised form and have been calibrated to reproduce the observed global efficiency of star formation. Our simulation assumes a $\Lambda$CDM cosmology with the parameters $\Omega_m = 0.2726$, $\Omega_\Lambda = 0.7274$, $\Omega_b = 0.0456$, $\sigma_8 = 0.809$, $n_s = 0.963$, and $H_0 = 100\,h$ km s$^{-1}$ Mpc$^{-1}$ with $h = 0.704$. Initial conditions are generated at $z = 127$ in a periodic box with a side length of 75 $h^{-1}$ Mpc = 106.5 Mpc. The initial gas temperature at $z = 127$ is set to 245 K. We have also performed a second simulation, Illustris-Dark, which does not include baryons and the related feedback, but is otherwise identical to Illustris. Dark matter haloes were identified in an on-the-fly manner during the simulation for each snapshot using a friends-of-friends (FOF) algorithm with a linking length of 0.2 times the mean particle separation. The minimum dark matter particle number was set to 32 for the FOF identification. Non-dark matter particles are attached to these FOF primaries in a secondary linking stage[44]. Subsequently, gravitationally bound substructures are identified using the SUBFIND algorithm[44,45]. 16 million CPU hours were needed to evolve the simulation from the starting redshift $z = 127$ to $z = 0$, using 8,192 cores and an equal number of MPI-ranks. An additional 3 million CPU hours were spent on carrying out the on-the-fly galaxy identification with the SUBFIND algorithm.

**Acknowledgements**

VS acknowledges support by the DFG Research Centre SFB-881 "The Milky Way System" through project A1, and by the European Research Council under ERC-StG EXAGAL-308037. GS acknowledges support from the HST grants program, number HST-AR-12856.01-A. Support for program #12856 was provided by NASA through a grant from the Space Telescope Science Institute, which is operated by the Association of Universities for Research in Astronomy, Inc., under NASA contract NAS 5-26555. LH acknowledges support from NASA grant NNX12AC67G and NSF grant AST-1312095. DX acknowledges support from the Alexander von Humboldt Foundation. SB is supported by NSF grant AST-0907969. The Illustris simulation was run on the CURIE supercomputer at CEA/France as part of PRACE project RA0844, and the SuperMUC computer at the Leibniz Computing Centre, Germany, as part of GCS-project pr85je. Further simulations were run on the Harvard Odyssey and CfA/ITC clusters, the Ranger and Stampede supercomputers at the Texas Advanced Computing Center through XSEDE, and the Kraken supercomputer at Oak Rridge National Laboratory through XSEDE. Fig. 1b is based on observations made with the NASA/ESA Hubble Space Telescope. These data were obtained from the Mikulski Archive for Space Telescopes (MAST) at the Space Telescope Science Institute (STScI). These observations were associated with programs 9352, 9425, 9488, 9575, 9793, 9978, 10086, 10189, 10258, 10340, 10530, 11359, 11563, 12060, 12061, 12062, 12099, and 12177, and compiled for the Hubble eXtreme Deep Field data release version 1.0 (http://archive.stsci.edu/prepds/xdf/). Support for MAST for non- HST data is provided by the NASA Office of Space Science via grant NNX13AC07G and by other grants and contracts.

**Author Contributions**

M.V., L.H., D.S., V.S., and S.G. conceived and planned the project. M.V., S.G., D.S., and P.T. developed the galaxy formation model. V.S. developed the AREPO code. M.V. generated initial conditions. V.S., M.V., and S.G. ran the simulations. M.V. performed the main analysis. G.S. and P.T. constructed the mock images. S.B. provided statistics of the inter-galactic medium. D.X. and D.N. provided post-processing tools. M.V., S.G., V.S., P.T., and L.H. interpreted the results. M.V. and S.G. wrote the manuscript with contributions from co-authors.

**Author Information**

Reprints and permissions information is available at www.nature.com/reprints. The authors declare no competing financial interests. Readers are welcome to comment on the online version of the paper. Correspondence and requests for materials should be addressed to M.V. (mvogelsb@mit.edu)

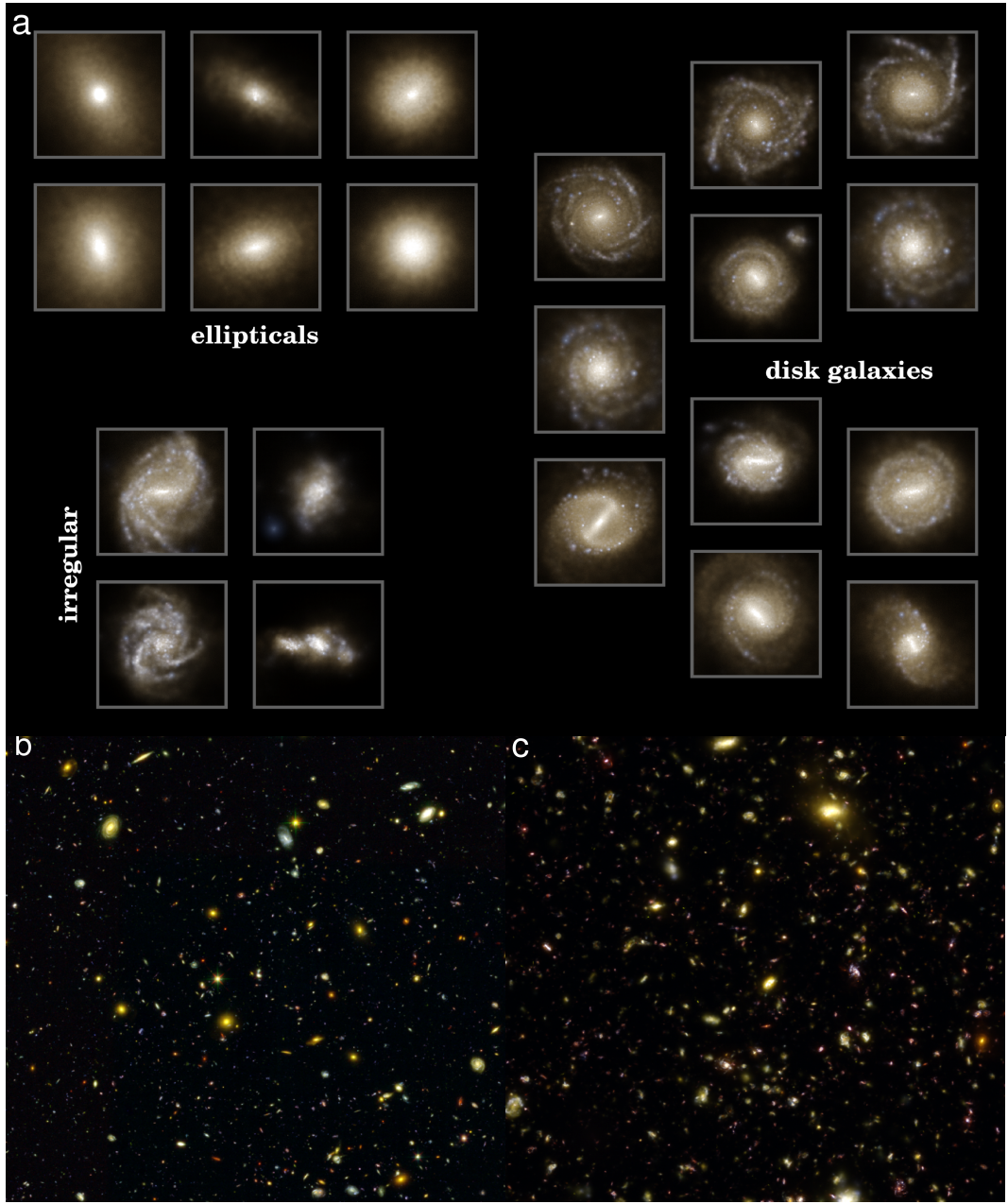

**Figure 1: Mock images of the simulated galaxy population. a,** Stellar light distributions (g,r,i bands) for a sample of galaxies at z = 0 arranged along the classical Hubble sequence for morphological classification. Our simulation produces a variety of galaxy types ranging from ellipticals to disk galaxies to irregular systems, the latter mostly resulting from interactions and mergers. **b,** HST UDF image (2.8 arcmin on a side) in B, Z, H bands convolved with Gaussian point-spread functions of σ = 0.04, 0.08, and 0.16 arcsec, respectively. **c,** HST mock observation from Illustris.

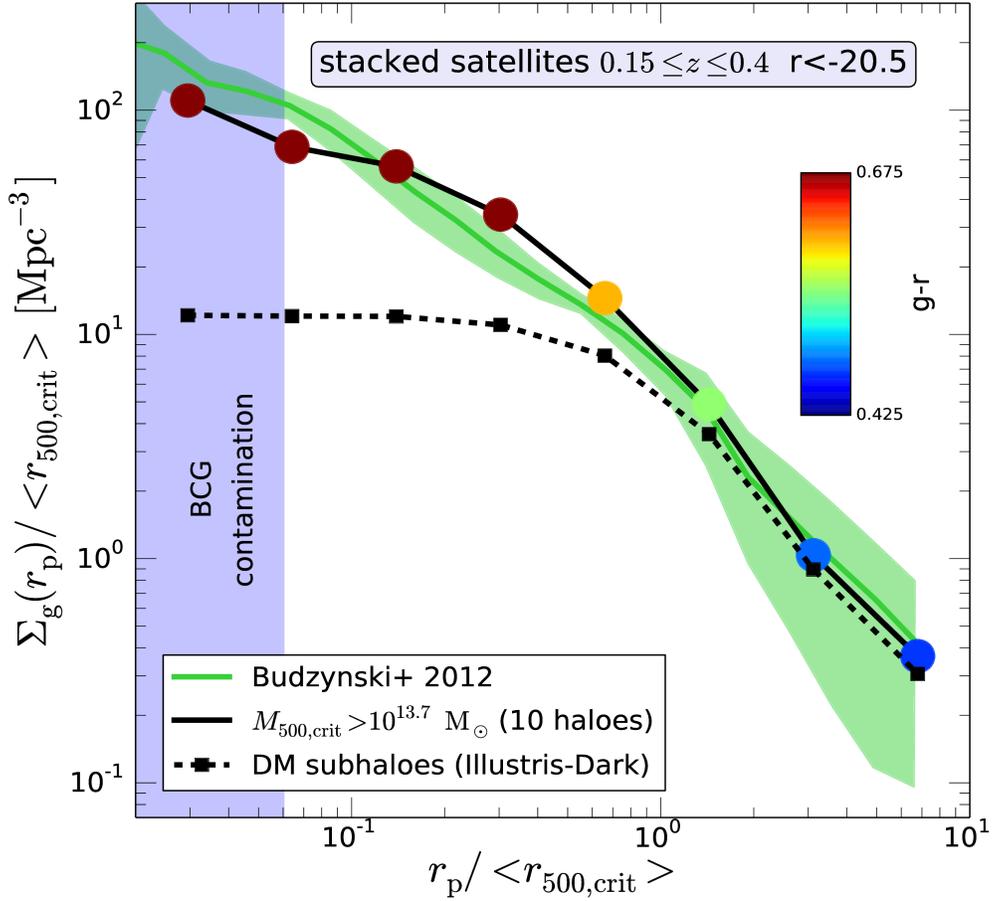

**Figure 2: Projected number density profile of satellite galaxies in galaxy clusters.** Symbols show a simulation sample as a function of cluster-centric projected radius, considering 10 clusters more massive than $M_{500,crit} = 10^{13.7}$ $M_\odot$ and all their satellite galaxies brighter than r < −20.5 within 3 × $R_{500,crit}$ along the line of sight from the cluster center. The symbol colour reflects the average (g-r) colour for galaxies at that distance. A reddening towards the cluster center is clearly visible. Observational data[13] is shown for comparison (green line), while the dashed line gives the subhalo profile of the corresponding dark matter-only simulation (normalised to match the satellite profile at large radii). The left blue region marks the area of incompleteness in the observed number density profiles due to obscuration from the brightest cluster galaxy. The green shaded region marks the observational uncertainties: outside of the region of contamination of the brightest cluster galaxy (BCG) the scatter within the observational stack dominates, whereas Poisson errors are shown within that region.

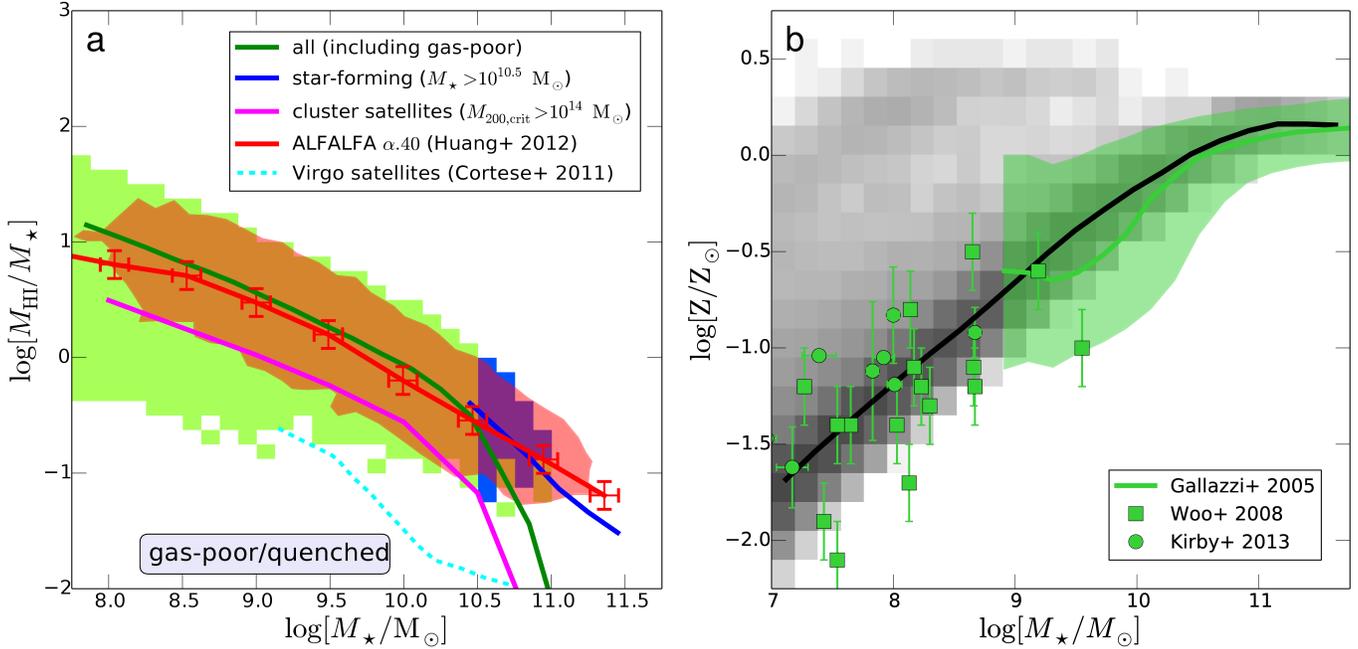

**Figure 3: Neutral hydrogen and metal content of galaxies as a function of their stellar mass. a,** Predicted HI richness of galaxies compared to the α.40 sample of the ALFALFA survey. Filled regions mark bins with more than 20 galaxies (red: ALFALFA, green: all galaxies, blue: star-forming galaxies). Error bars indicate s.e.m. for individual galaxies. A separate HI richness relation for the satellites of the most massive cluster is included and compared to HI measurements of Virgo satellites[20] to demonstrate the strong impact of environment on the H I content of satellites. **b,** Stellar metal content of simulated galaxies (black line and grey shaded histogram) compared to observations[24–26] (green). Reassuringly, the trend of the median relation including the flatting above M ~ $10^{11}$ M$_\odot$ is well recovered. The green shaded region represents the s.d. The error bars represent the s.e.m. for individual galaxies.

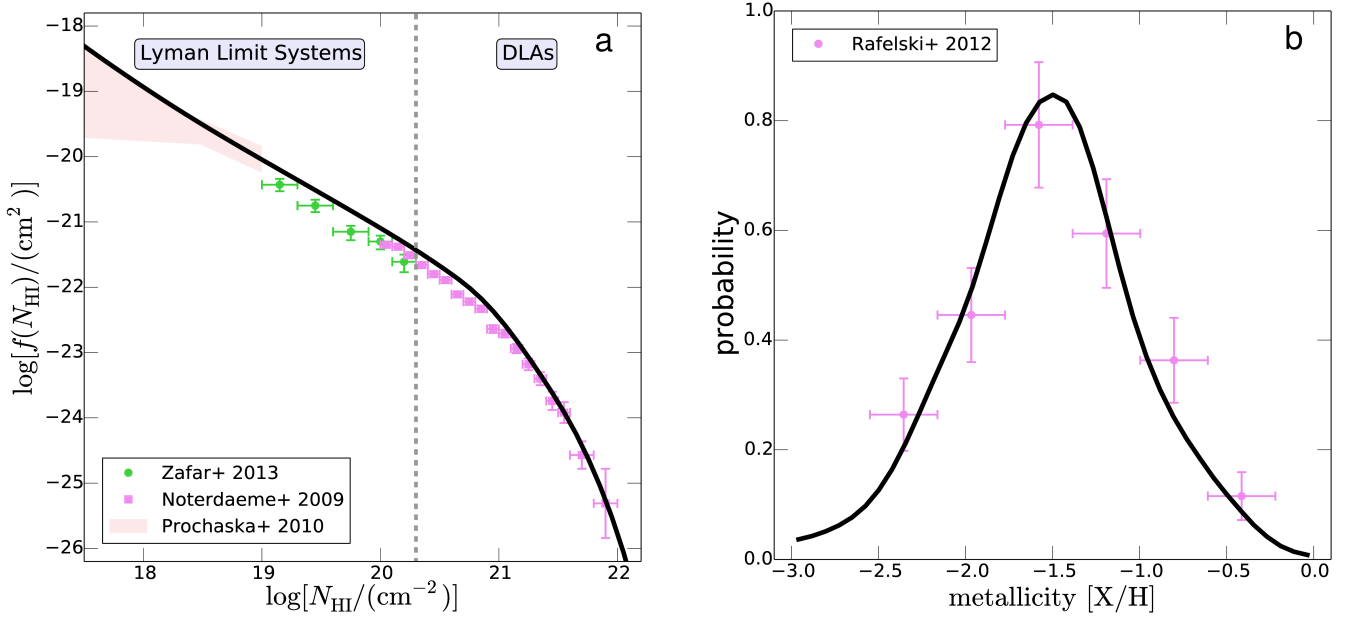

**Figure 4: Large-scale characteristics of neutral hydrogen. a,** Column density distribution function (CDDF) of neutral hydrogen at z = 3 compared to observations[27, 28,46]. The dashed vertical line shows the density threshold separating Damped Lyman-α systems (DLAs) and Lyman Limit systems. The shaded region shows an estimate for the CDDF constrained by the assumption of a power law fit and the observed incidence of Lyman Limit Systems. The vertical error bars represent the s.e.m, and the horizontal ones represent the binning. **b,** Probability density function of the DLA metallicity in units of solar metallicity at z = 3 is compared to observational findings[32]. The observational error bars show the s.e.m. derived from the number of observed spectra in each metallicity bin. Bin widths have been chosen to be larger than the maximal uncertainty in each individual metallicity measurement.

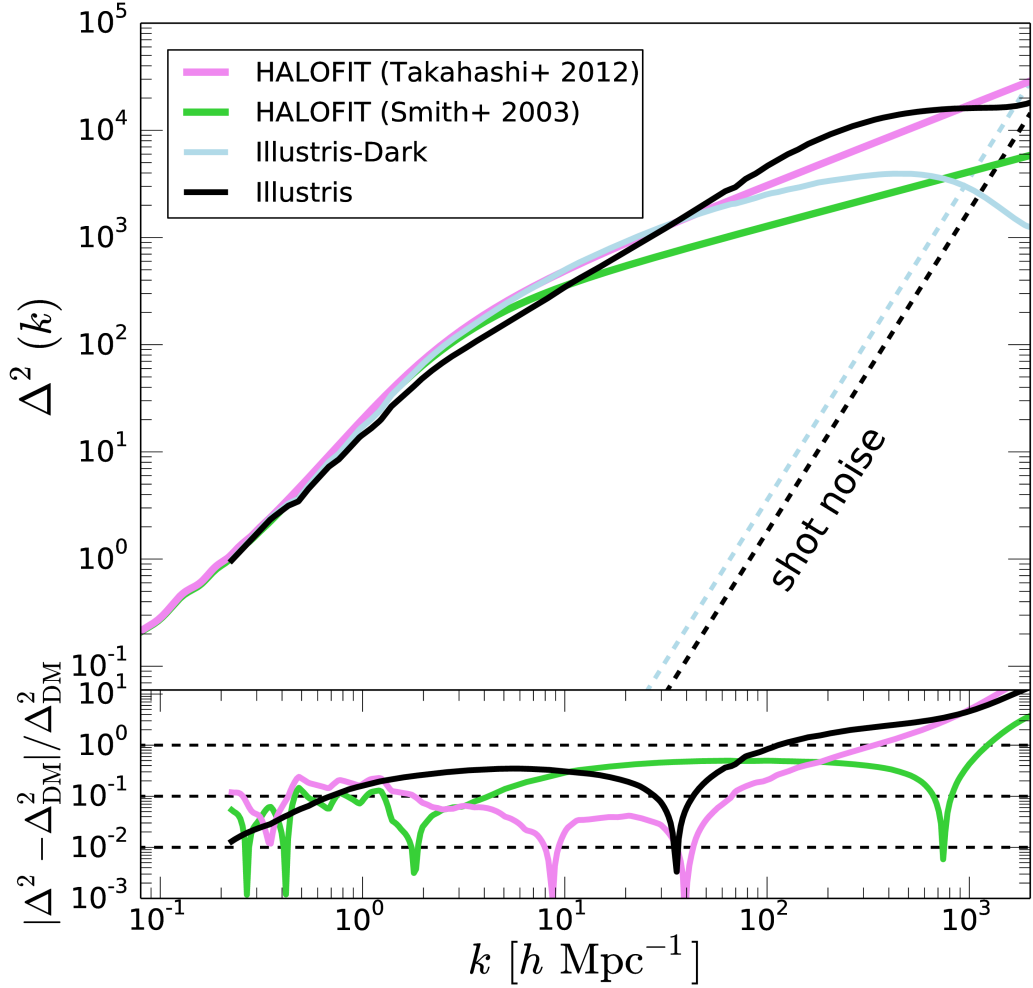

**Figure 5: Non-linear matter power spectrum.** The dimensionless total matter power spectrum, $\Delta^2(k)$, of the Illustris simulation (top panel, black line) differs significantly, due to baryonic effects, from that of the dark matter-only counterpart Illustris-Dark (light blue). Analytic fitting models[39,40] (green and pink) do not provide an adequate description of the hydrodynamic results. The lower panel shows their relative difference, highlighting that baryonic effects exceed 1% already on scales smaller than $k \sim 1\ h$ Mpc$^{-1}$. The theoretical shot noise level (shown as thin dashed lines) has been subtracted in the measurements.

# Methods

**Simulation code**

Our simulation code, AREPO, uses an unstructured Voronoi tessellation of the simulation volume, where the mesh-generating points of this tessellation are moved with the gas flow. The adaptive mesh is used to solve the equations of ideal hydrodynamics with a finite volume approach using a second-order unsplit Godunov scheme with an exact Riemann solver. This approach is under most circumstances superior to traditional smoothed particle hydrodynamics (SPH), and also to Eulerian adaptive mesh refinement (AMR)[47,48,51,52]. This scheme naturally produces extended disk galaxies without invoking extreme forms of stellar feedback or star formation[53], which was a major problem of previous galaxy formation simulations. The gravity calculation employs a Tree-PM scheme[54], where long-range forces are determined with a particle-mesh method (PM) while short-range forces are computed via a hierarchical tree algorithm[55].

**Galaxy formation physics**

Our simulation accounts for a variety of astrophysical processes known to be relevant for galaxy formation[43].

Gas cooling rates are calculated as a function of gas density, temperature, metallicity, the radiation fields of active galactic nuclei and the spatially uniform but time-dependent ionising background radiation from galaxies and quasars[56], which completes HI reionisation at a redshift of z~6. We use a self-consistent calculation of the primordial cooling and complement it with the cooling contribution of the metals, based on pre-calculated cooling rate tables using CLOUDY[57]. All cooling rates include self-shielding corrections[58].

We employ a sub-resolution model of the interstellar medium to achieve numerical closure below our resolution scale. Gas with hydrogen number density above 0.13 cm$^{-3}$ follows an effective equation of state with a stochastic prescription for star formation following the Kennicutt-Schmidt law[59] and adopting a Chabrier initial mass function[60]. The effective equation of state assumes that the interstellar medium has a two-phase structure that is predominantly composed of cold clouds embedded in a tenuous, supernova-heated phase[61].

Once stellar populations are born, they can lose mass, for example through stellar winds or supernovae. This mass is returned to the gas phase and enriches the gas surrounding stellar populations. We track the evolution of stars and model supernovae of type Ia, type II, and the asymptotic giant branch phases of stars. We trace the evolution of nine elements in total (H, He, C, N, O, Ne, Mg, Si, Fe), each advected as a passive scalar.

Stellar feedback is realised through a kinetic wind scheme with a velocity scaling based on the local one-dimensional dark matter velocity dispersion ($3.7\sigma_{DM,1D}$), and a mass loading inferred from energy conservation assuming $1.09 \times 10^{51}$ erg per SNII. We use a sub-grid metal-loading scheme that regulates the degree of wind enrichment such that 40% of the local interstellar medium metals are ejected by supernova-driven galactic winds. This is required to simultaneously reproduce the stellar mass content of low mass haloes and their gas oxygen abundances[62].

We include procedures for supermassive black hole (SMBH) seeding, accretion and merging[49]. Feedback from SMBHs operates in either a quasar-mode or a radio-mode, depending on their accretion rate[50]. In addition, a prescription for radiative SMBH feedback is included that modifies the ionisation state and hence the net cooling rate of nearby gas. SMBHs are seeded in friends-of-friends groups more massive than $7.1 \times 10^{10}$ $M_\odot$ with a seed mass of $1.4 \times 10^5$ $M_\odot$.

The free parameters of our model are set to physically plausible values and have been adjusted within the allowed range to roughly reproduce the relation between mean stellar mass and halo mass inferred from abundance matching analysis. The resulting parameter settings have been tested on smaller scale simulations[43] and high-resolution zoom-in simulations of individual Milky Way-like haloes[63].

**Initial conditions**

To create initial conditions we use the Boltzmann code CAMB[64,65] to compute the linear power spectrum of a ΛCDM cosmology with the parameters $\Omega_m = 0.2726$, $\Omega_\Lambda = 0.7274$, $\Omega_b = 0.0456$, $\sigma_8 = 0.809$, $n_s = 0.963$, and $H_0 = 100\ h$ km s$^{-1}$ Mpc$^{-1}$ with $h = 0.704$. These parameters are consistent with the latest Wilkinson Microwave Anisotropy Probe (WMAP)-9 measurements[66], but slightly offset from the first year results of the Planck mission[1]. However, a recent re-analysis of the Planck data found parameters more consistent with pre-Planck cosmic microwave background analyses and astronomical observations[67].

We create a random realisation of this cosmology in periodic boxes with a side length of 75 $h^{-1}$ Mpc ≈ 106.5 Mpc, starting from an initial "glass-like" particle configuration[68] composed of one thousand $182^3$ particle tiles. We employ a $3{,}640^3$ Fast Fourier Transform to calculate the displacement field and use Lagrangian perturbation theory (Zel'dovich approximation[69]) to displace particles, and we de-convolve the input power spectrum for smoothing effects due to the interpolation off this grid. Initial conditions are generated at z = 127 with mesh-generating points added to the initial conditions by splitting each original particle into a dark matter and gas cell pair, displacing them with respect to one another such that two interleaved grids are formed, keeping the centre-of-mass of each pair fixed. The initial gas temperature at z = 127 is set to 245 K based on a RECFAST[70,71] calculation. We have generated 100 different random fields and inspected their power spectra and mass functions at z = 0 to make sure that we do not simulate an unusual or extreme density field that is dominated by a few large clusters or voids due to cosmic variance.

**Simulation details**

The simulation volume contains initially 6,028,568,000 hydrodynamic cells and the same number of dark matter particles resulting in a dark matter mass resolution of 6.26 × $10^6$ M$_\odot$, and a baryonic mass resolution of $1.26 \times 10^6$ M$_\odot$. The gravitational softening length of dark matter particles is fixed in comoving coordinates ($\varepsilon_{DM} = 1\ h^{-1}$ kpc). For baryonic particles (stars and SMBHs), we limit the softening length to a maximum physical scale ($\varepsilon_{baryon} = 0.5\ h^{-1}$ kpc). Gas cells use an adaptive softening length tied to their cell radius with a floor given by the softening length of the collisionless baryonic particles. We employ a (de-) refinement scheme which keeps the cell masses typically within a factor of two of a specified target mass set to $1.26 \times 10^6$ M$_\odot$, and a regularisation scheme steering the mesh towards a computationally efficient centroidal configuration[5,43,47]. The smallest cells in Illustris have a typical extent of 48 pc. For the least massive cells we achieve a mass resolution of $1.5 \times 10^4$ M$_\odot$.

**Galaxy identification**

We derive the various galaxy properties from the gravitationally bound mass that is contained within a radius r$_*$ that equals twice the stellar half mass radius of each SUBFIND (sub)halo. Using this definition, the galactic stellar mass does not significantly differ from the total stellar mass for low mass systems, but some of the intra-cluster light for massive systems is excluded. We have checked this definition against surface brightness cuts in different bands and find it to give similar results as such more elaborate methods for excluding intra-cluster light. We have extended this radius by 50% for measuring the galactic HI masses. Stellar population synthesis models[72] were used to associate the stars in our simulation with observable broad band luminosities. Producing stellar images of galaxies requires assigning colour values to specific bands. Specifically, we make an RGB mapping of the (g,r,i) bands using a commonly employed asinh scaling[73].